\documentclass[camera]{jpaper}

\usepackage[nocompress]{cite}
\usepackage{array}

\makeatletter
\let\MYcaption\@makecaption
\makeatother

\usepackage[font=footnotesize]{subcaption}

\makeatletter
\let\@makecaption\MYcaption
\makeatother

\usepackage{fixltx2e}
\usepackage{dblfloatfix}
\usepackage[nolessnomore, italic]{mathastext}
\usepackage[T1]{fontenc}
\usepackage[usenames,dvipsnames,svgnames,table]{xcolor}
\usepackage[normalem]{ulem}
\usepackage{enumitem}
\usepackage{setspace}
\usepackage{indentfirst}
\usepackage{fancyhdr}
\usepackage{authblk}
\usepackage{url}
\usepackage[binary-units=true]{siunitx}
\usepackage[us,12hr]{datetime}
\usepackage{flushend}
\usepackage[hidelinks]{hyperref}

\usepackage{xspace}
\usepackage{mystyle}
\usepackage{algpseudocode}
\usepackage{color,soul}
\definecolor{pastelyellow}{rgb}{0.99, 0.99, 0.59}

\newif\ifcameraready
\camerareadytrue

\newcommand{\versionnum}[0]{4}

\newcommand{\X}[0]{Zorua\xspace}

\newcommand{\One}{\emph{(i)}~}
\newcommand{\two}{\emph{(ii)}~}
\newcommand{\three}{\emph{(iii)}~}
\newcommand{\PrintTodo}{true}
\newcommand{\todo}[2]{
\ifx\PrintTodo\undefined
\textcolor{red}{#2\xspace}
\else
\textcolor{red}{TODO[#1]:} \textcolor{blue}{\textit{#2}\xspace}
\fi
}

\ifcameraready
  \newcommand{\highlight}[1]{{\leavevmode{#1}}}
\else
  \newcommand{\highlight}[1]{{\leavevmode\color{PineGreen}{#1}}}
\fi

\usepackage[font={small,bf}]{caption}
\usepackage{caption}
\usepackage{subcaption}
\newcommand{\ignore}[1]{}
\newcommand{\squishlist}{
   \begin{list}{$\bullet$}
    { \setlength{\itemsep}{0pt}      \setlength{\parsep}{3pt}
      \setlength{\topsep}{3pt}       \setlength{\partopsep}{0pt}
      \setlength{\leftmargin}{1.5em} \setlength{\labelwidth}{1em}
      \setlength{\labelsep}{0.5em} } }

\newcommand{\squishlisttwo}{
   \begin{list}{$\numbered$}
    { \setlength{\itemsep}{0pt}    \setlength{\parsep}{0pt}
      \setlength{\topsep}{0pt}     \setlength{\partopsep}{0pt}
      \setlength{\leftmargin}{2em} \setlength{\labelwidth}{1.5em}
      \setlength{\labelsep}{0.5em} } }

\newcommand{\squishend}{
    \end{list}  }

\begin{document}
\title{\Large{Decoupling GPU Programming Models from Resource Management for \\Enhanced Programming Ease, Portability, and Performance\\}}
\author{Nandita Vijaykumar$^1$\qquad Kevin Hsieh$^1$\qquad Gennady
Pekhimenko$^{2,3,1}$\qquad Samira
Khan$^4$
\vspace{2pt}\\
Ashish Shrestha$^{5,1}$\qquad Saugata Ghose$^1$\qquad Adwait
Jog$^6$\qquad
Phillip B. Gibbons$^1$\qquad Onur Mutlu$^{7,1}$}
\affil{
\emph{$^1$Carnegie Mellon University}
\qquad\emph{$^2$University of Toronto}\qquad
\emph{$^3$Microsoft Research}%
\vspace{2pt}\\
\emph{$^4$University of Virginia}\qquad%
\emph{$^5$Intel}\qquad%
\emph{$^6$College of William and Mary}\qquad%
\emph{$^7$ETH Z{\"u}rich}
}

\date{}
\maketitle
 
\begin{abstract}
This paper summarizes the idea of Zorua, which was published in MICRO 2016~\cite{zorua}, 
and examines the work's significance and future potential.
The application resource specification{\textemdash}a static specification of several parameters
such as the number of threads and the scratchpad memory usage per thread
block{\textemdash}forms a critical component of modern GPU programming models.
This specification determines the parallelism, and hence performance, of the
application during execution because the corresponding on-chip hardware resources are allocated and managed based
on this specification. This \emph{tight-coupling} between the software-provided
resource specification and resource management in hardware leads to
significant challenges in programming ease, portability, and performance. Zorua is a new resource virtualization framework,
that \emph{decouples} the programmer-specified resource usage of a
GPU application from the actual allocation in the on-chip hardware
resources. 
\X enables this decoupling by \emph{virtualizing} each
resource transparently to the programmer.  

The virtualization provided
by \X builds on two key concepts{\textemdash}\emph{dynamic allocation}
of the on-chip resources, and their \emph{oversubscription} using a
swap space in memory. \X provides a holistic GPU resource virtualization strategy designed
to {\One}adaptively \emph{control the extent} of oversubscription, and
\two \emph{coordinate} the dynamic management of multiple on-chip resources to maximize the
effectiveness of virtualization. We demonstrate that by providing the illusion of more resources than
physically available via controlled and coordinated virtualization, \X
offers several important benefits: \One \textbf{Programming Ease.} \X
eases the burden on the programmer to provide code that is tuned to
efficiently utilize the physically available on-chip resources. \two
\textbf{Portability.} \X alleviates the necessity of re-tuning an
application's resource usage when porting the application across GPU
generations. \three \textbf{Performance.} By dynamically allocating
resources and carefully oversubscribing them when necessary, \X
improves or retains the performance of applications that are already
highly tuned to best utilize the resources. The holistic
virtualization provided by \X has many other potential uses, e.g.,
fine-grained resource sharing among multiple kernels, low-latency 
preemption of GPU programs, and support for dynamic parallelism. 
\end{abstract} 

\section{Motivation: Key Challenges in \\ Modern GPUs}
Modern Graphics Processing Units (GPUs) offer high performance
and energy efficiency for many classes of applications
by concurrently executing thousands of threads. In order to execute,
each thread requires several major on-chip resources: \One registers,
\two scratchpad memory (if used in the program), and
\three a thread slot in the thread scheduler that keeps all the
bookkeeping information required for execution.

Today, these resources are {\em statically} allocated to threads based on several parameters{\textemdash}the number of threads
per thread block, register usage per thread, and scratchpad usage per
block. We refer to these static application parameters as the
\emph{resource specification} of the application. This resource specification
forms a critical component of modern GPU programming models (e.g.,
CUDA~\cite{cuda},
OpenCL~\cite{opencl}). The static
allocation over a fixed set of hardware resources based on the
software-specified resource specification creates a
\emph{tight coupling} between the program 
and the physical hardware resources. As a result of this tight coupling, for
each application, there are only a few optimized resource specifications that
maximize resource utilization. Picking a suboptimal specification
leads to underutilization of resources and hence, very often,
performance degradation. This leads to three key
difficulties related to obtaining good performance on modern GPUs: programming
ease, portability, and performance degradation.  

\begin{figure}[t]
\centering
\includegraphics[width=0.45\textwidth]{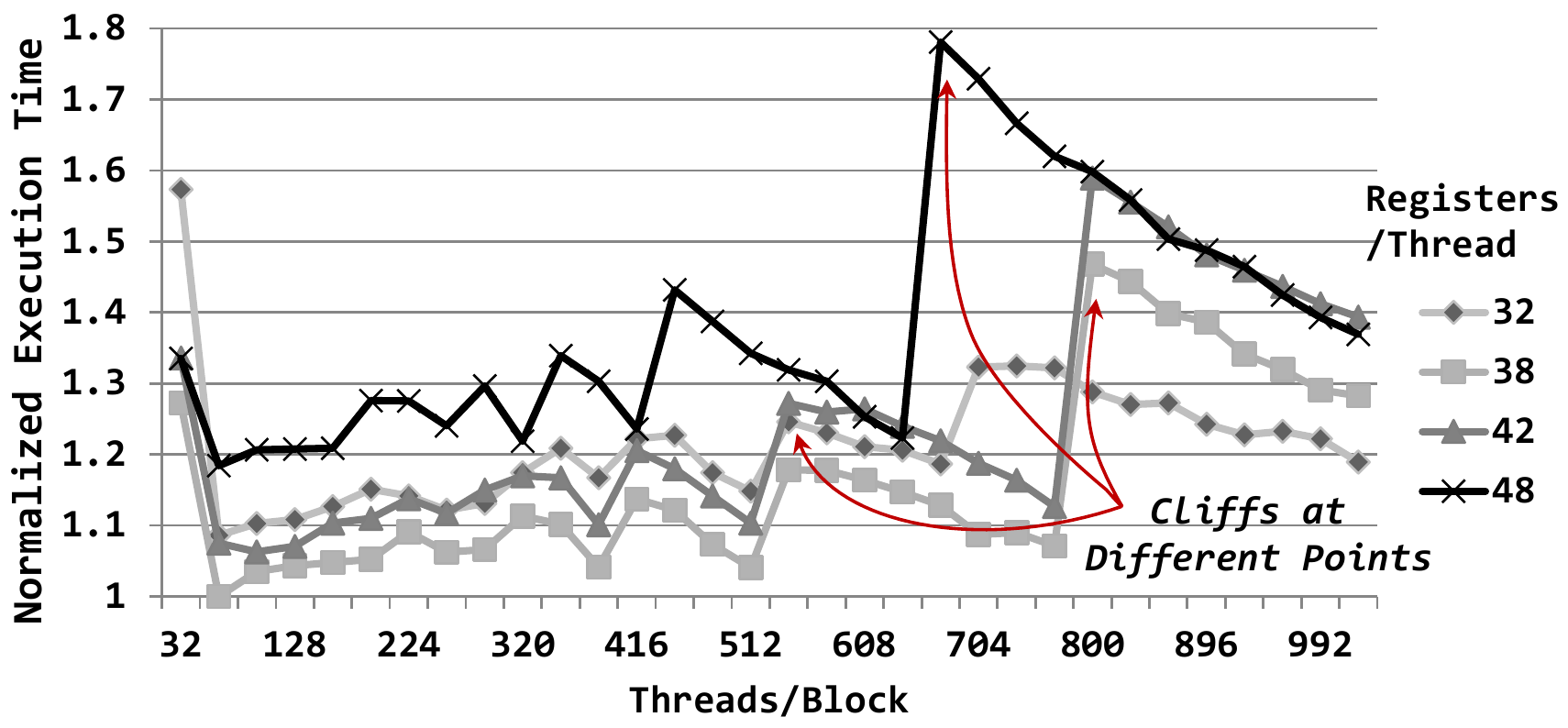} 
\caption{Performance cliffs in \emph{Minimum Spanning Tree} (\emph{MST}) when
run on the NVIDIA GTX 745.
Reproduced from~\cite{zorua}.} 
\label{fig:mst-cliff}
\end{figure}

\textbf{Programming Ease.}
First, the burden falls upon the
programmer to optimize the resource specification.
For a naive programmer, this is a challenging
task because, in addition to selecting a specification suited to an
algorithm, the programmer needs to be aware of the details of the GPU architecture
to fit the specification to the underlying hardware resources. This
\emph{tuning} is easy to get wrong because there are \emph{many}
highly suboptimal performance points in the specification space, and
even a minor deviation from an optimized specification can lead to a
drastic drop in performance due to lost parallelism. We refer to such
drops as \emph{performance cliffs}. 
Even a
small change in one resource can result in a significant
performance cliff, degrading performance by as much as 50\%.
Figure~\ref{fig:mst-cliff} depicts multiple sizable cliffs in an example
application, when different resource specifications are used when the program
is run on a real modern GPU, the NVIDIA GTX 745.\footnote{Our MICRO 2016
paper~\cite{zorua} describes the experimental methodology for collecting these real system
results.}

\begin{figure}[h]  
\centering 
\includegraphics[width=0.40\textwidth]{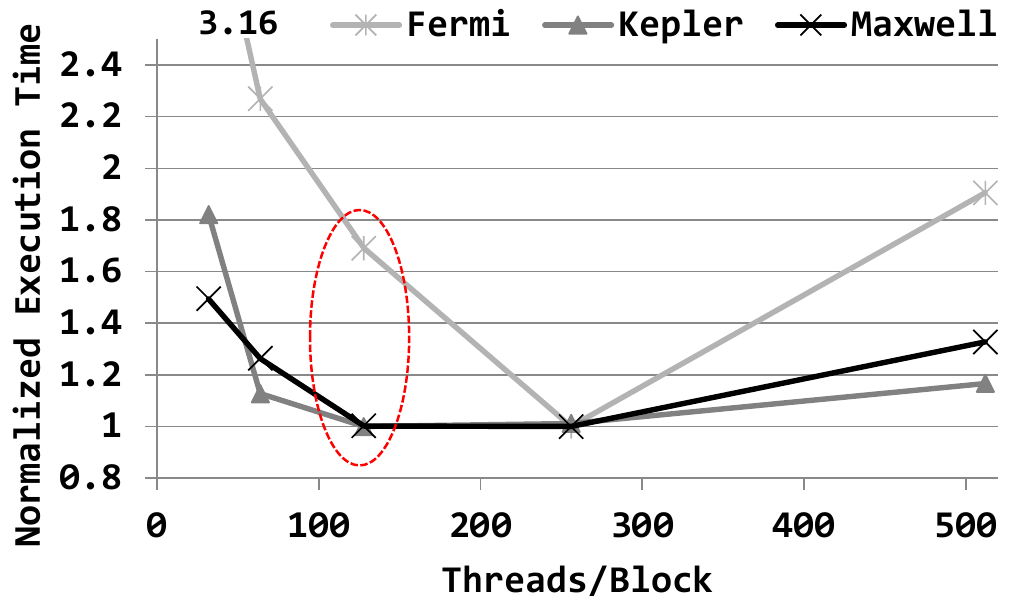}
\caption{Performance
variation across different GPU generations from NVIDIA (Fermi, Kepler, and
Maxwell) for 
\emph{Discrete Fourier Transform (DCT)}. Reproduced from~\cite{zorua}.}
\label{fig:port} 
\end{figure}

\textbf{Portability.}
Second, different GPUs have varying quantities
of each of the resources. Hence, an optimized specification on one GPU
may be highly suboptimal on another. This lack of \emph{portability} necessitates
that the programmer \emph{re-tune} the resource specification of the application for
\emph{every} new GPU generation. This
problem is especially significant in virtualized environments, such as data
centers, cloud computing, or compute
clusters, where the same program may run on a wide range of GPU architectures.
Figure~\ref{fig:port} depicts the 69\% performance loss when porting
optimized code from the NVIDIA
Kepler~\cite{kepler}/Maxwell~\cite{maxwell} architectures to the NVIDIA
Fermi~\cite{fermi} architecture.

\textbf{Performance.}
Third, for a programmer who chooses to employ software
optimization tools (e.g.,
auto-tuners~\cite{toward-autotuning,atune,maestro,parameter-profiler,autotuner1,autotuner-fft}) or manually tailor the program to fit the
hardware,
performance is still constrained by the \emph{fixed, static} resource
specification. It is well
known~\cite{virtual-register,compiler-register,shmem-multiplexing, caba,
kayiran-pact16, largewarp}
that the on-chip resource requirements of a GPU application vary throughout
execution. Since the program (even after auto-tuning) has to {\em statically}
specify its {\em worst-case} resource requirements, severe
\emph{dynamic underutilization} of several GPU
resources ensues~\cite{caba},
leading to suboptimal performance.

\section{A Holistic Approach to\\ Resource Virtualization}

To address these three challenges at the same time, we propose Zorua, a new framework that
\emph{decouples} an application's resource specification from the available hardware
resources by \emph{virtualizing} all three major resources (i.e., scratchpad
memory, register file, and thread slots) in a holistic manner. This
virtualization provides the illusion of more resources to the GPU programmer and
software than physically available, and enables the runtime system and the
hardware to {\em dynamically} manage multiple resources in a manner that is transparent
to the programmer.

\subsection{Key Concepts}
The virtualization strategy used by \X is
built upon two key concepts. First, to mitigate performance cliffs when we do
not have enough physical resources, we \emph{oversubscribe} resources by a small
amount at runtime, by leveraging their dynamic underutilization and maintaining a
swap space (in main memory) for the extra resources required. Second, \X
improves utilization by determining the runtime resource requirements of an
application. It then allocates and deallocates resources dynamically, managing
them \One \emph{independently} of each other to maximize each resource's utilization; and \two in a \emph{coordinated} manner, to enable efficient execution of each
thread with all its required resources available.

\begin{figure}[h] \centering
\includegraphics[width=0.49\textwidth]{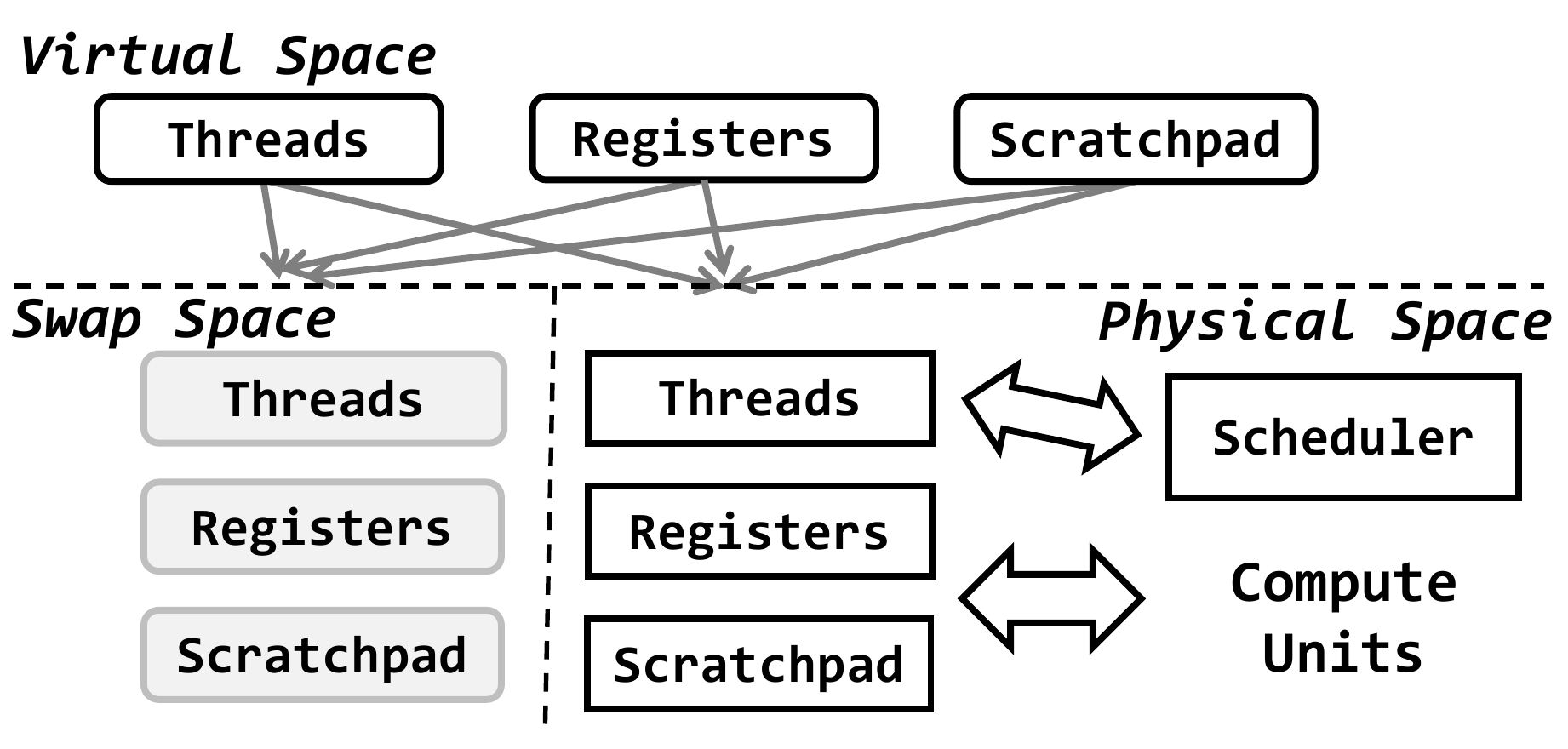}
\caption{High-level overview of \X. Reproduced from \cite{zorua}.} \label{fig:overview} \end{figure}

Figure~\ref{fig:overview} depicts the high-level overview of the virtualization
provided by \X. 
The \emph{virtual space} refers to the \emph{illusion} of the quantity of available
resources. The \emph{physical space} refers to the \emph{actual} hardware resources
(specific to the target GPU architecture), and the \emph{swap space} refers to the resources
that do \emph{not} fit in the physical space and hence are \emph{spilled} to other
physical locations.
For the register file and scratchpad memory, the swap space is mapped to the global
memory space in the memory hierarchy. For threads, only
those that are mapped to the physical space are available for scheduling
and execution at any given time. If a thread is mapped to the swap space, its
state (e.g., the PC) is saved in memory. Resources in the
virtual space can be freely re-mapped between the physical and swap
spaces to maintain the illusion of the virtual space resources.
 
\subsection{Challenges in Virtualization}
Unfortunately, oversubscription means
that latency-critical resources, such as registers and scratchpad, may be
swapped to memory at the time of access, resulting in high overheads in performance
and energy. This leads to two critical challenges in designing a framework to
enable virtualization. The first challenge is to effectively determine the
\emph{extent} of virtualization, i.e., by how much each resource appears to be
larger than its real physical amount, such that we can
\emph{minimize} oversubscription while still reaping its benefits. This is difficult as
the resource requirements continually vary during runtime. The second challenge
is to minimize accesses to the swap space. This requires \emph{coordination} in
the virtualized management of \emph{multiple resources}, so that enough of each
resource is available
on-chip at the same time when needed.

\subsection{Design Ideas}
To solve these challenges, \X employs two key ideas. First, we leverage the software
(the compiler) to provide annotations with information regarding the future resource
requirements of each \emph{phase} of the application. This information enables the
framework to make intelligent dynamic decisions ahead of time, with respect to both 
the extent of oversubscription and the allocation/deallocation of resources.
Second, we use an adaptive runtime system to control the allocation of resources. This allows us to \One
dynamically alter the extent of
oversubscription; and \two continuously coordinate the allocation of multiple
on-chip resources and the mapping between their virtual and physical/swap
spaces; depending
on the varying runtime requirements of each thread. We briefly describe each design idea
in turn.
\subsubsection{Leveraging Software Annotations of Phase Characteristics}
\label{sec:key_idea_phases} 
We observe that the runtime variation in resource requirements
typically occurs at 
the granularity of \emph{phases} of a few tens of instructions. This variation
occurs because different parts of kernels perform different operations that
require different resources. For example, loops that primarily load/store data
from/to 
scratchpad memory tend to be less register heavy. Sections of
code that perform specific computations (e.g., matrix transformation, graph
manipulation), can either be register heavy or primarily operate out of
scratchpad. Often, scratchpad memory is used for only short
intervals~\cite{shmem-multiplexing}, e.g., when data exchange between threads is
required, such as for a reduction operation. 

\begin{figure}[h] \centering
\includegraphics[width=0.49\textwidth]{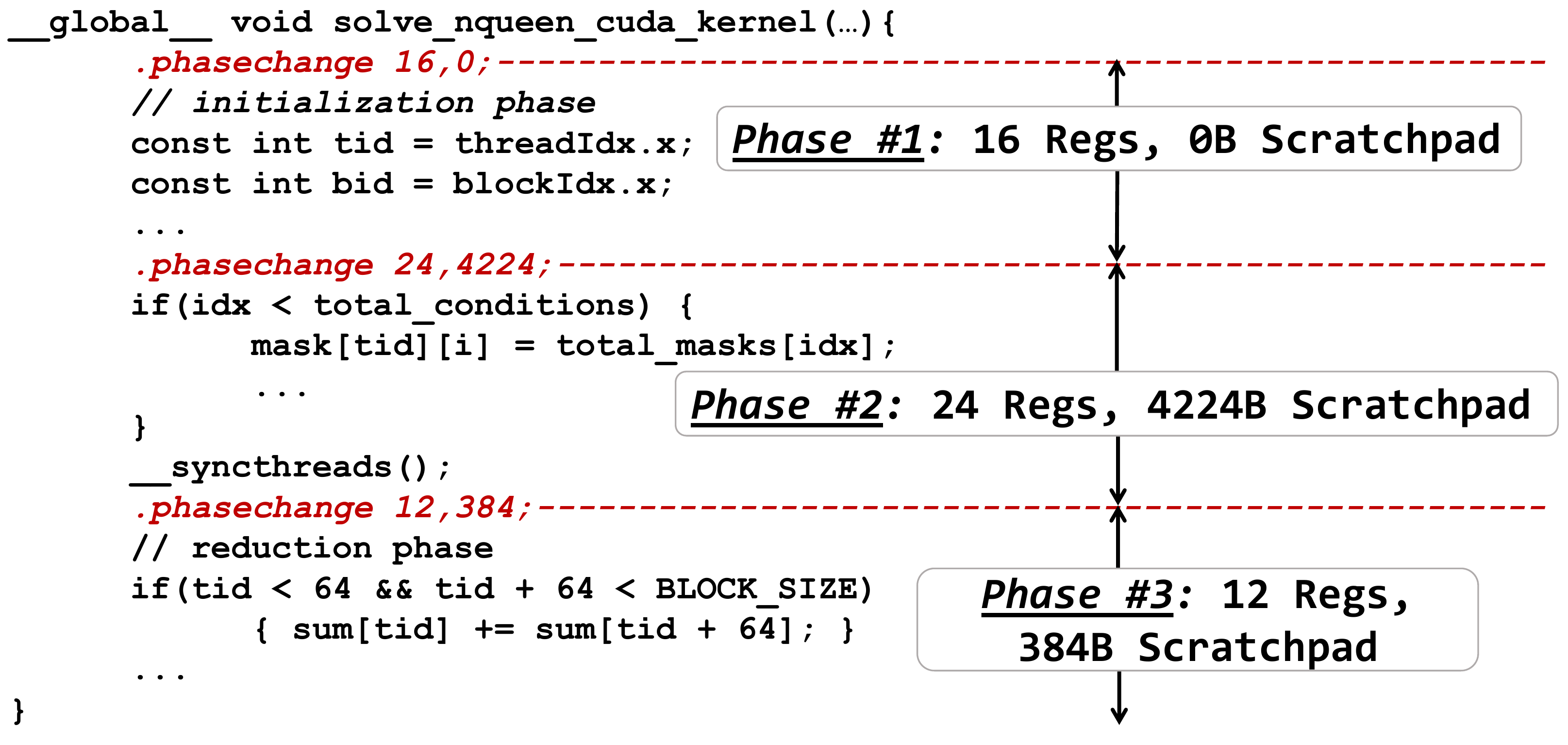}
\caption{Example phases from \emph{N-Queens Solver (NQU)}. Reproduced from~\cite{zorua}.} \label{fig:phases} \end{figure}

Figure~\ref{fig:phases} depicts a few example phases from the 
\emph{N-Queens
Solver (NQU)}~\cite{NQU} kernel. \emph{NQU} is a scratchpad-heavy application, but it does
not use the
scratchpad at all during the initial computation phase. During its second phase, it performs
its primary computation out of the scratchpad, using as much as 4224B. During its
last phase, the scratchpad is used only for reducing results, which requires
only 384B. There is also significant variation in the maximum
number of live registers in the different phases, as shown in
Figure~\ref{fig:phases}. 

In order to capture both the resource requirements as well as their variation
over time, we 
partition the program into a number of \emph{phases}. A phase is a
sequence of instructions with sufficiently different resource requirements than
adjacent phases.\footnote{We refer the reader to Section~4.6 of our MICRO 2016 paper~\cite{zorua} for
specific details on how phases are identified.} Barrier or fence operations also indicate a change in
requirements for a different reason{\textemdash}threads that are waiting at a barrier do
not immediately require the thread slot that they are holding. 
We interpret barriers and fences as phase boundaries since they potentially alter
the utilization of their thread slots. The compiler inserts special instructions
called \emph{phase specifiers} to mark the start of a new phase. Each phase
specifier contains information regarding the resource requirements of the next
phase. Phase changes are shown as ``\texttt{.phasechange}'' pragmas in
Figure~\ref{fig:phases}. 
 
A phase forms the basic unit for resource allocation and
deallocation, as well as for making oversubscription decisions. It offers
a finer granularity than an \emph{entire thread} to make such decisions.
The phase specifiers provide information on the \emph{future resource usage} of
the thread at a phase boundary. This enables \One preemptively controlling the
extent of oversubscription at runtime, and \two dynamically allocating and
deallocating resources at phase boundaries to maximize utilization of the
physical resources.

\subsubsection{Control with an Adaptive Runtime System}
\label{sec:key_idea_coordinator}
\begin{figure}[h] \centering
\includegraphics[width=0.45\textwidth,scale=0.8]{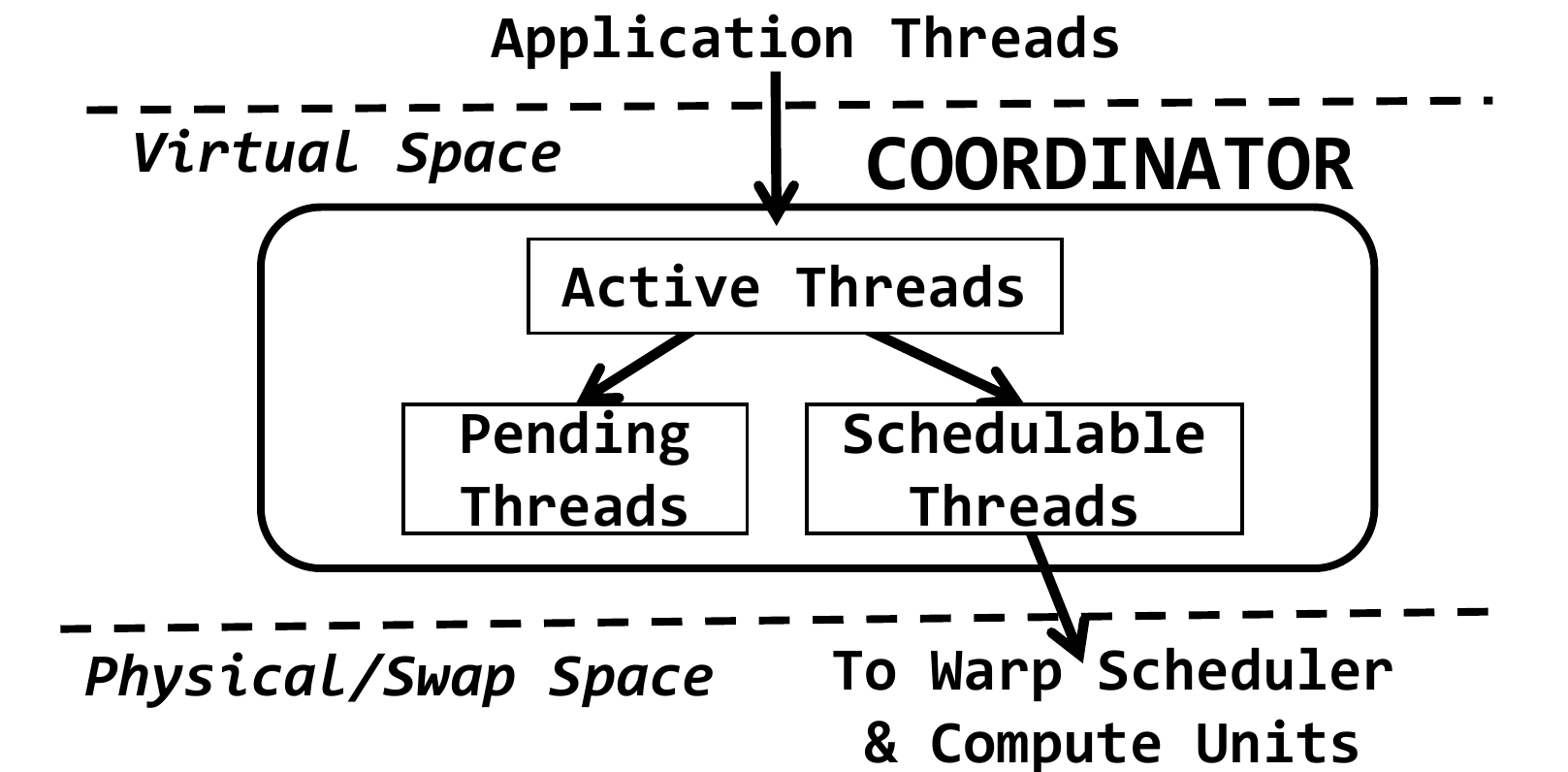}
\caption{Overview of the coordinator. Reproduced from~\cite{zorua}.} \label{fig:coordinator} \end{figure}

Phase specifiers provide information to make oversubscription and allocation/deallocation
decisions. However, we still need a way to make decisions on the
extent of oversubscription and appropriately allocate resources at runtime. To this
end, we use an adaptive runtime system, which we refer to as the
\emph{coordinator}. Figure~\ref{fig:coordinator} presents an overview of the
coordinator.

The virtual space enables the illusion of a larger amount of each of the
resources than what is physically available, to adapt to different application
requirements. This illusion enables higher thread-level parallelism than what
can be achieved with solely the fixed, physically available resources, by
allowing more threads to execute concurrently. The size of the virtual space at a given
time determines this parallelism, and those threads that are effectively
executed
in parallel are referred to as \emph{active threads}. All active threads have
thread slots allocated to them in the virtual space (and hence can be executed),
but some of
them may \emph{not} be mapped to
the physical space at any given time. As discussed previously,
the resource requirements of each application continuously change during
execution. To adapt to these runtime changes, the coordinator leverages
information from the phase specifiers to make decisions on oversubscription. The
coordinator makes these decisions at
every phase boundary and thereby controls the size of the virtual space for each resource.

\subsection{Zorua: An Overview}
To address the challenges
in virtualization by leveraging the above ideas, \X employs a
software-hardware codesign that comprises three components: \One
\textbf{\emph{The compiler}} annotates the program by adding special
instructions (\emph{phase specifiers}) to partition it into \emph{phases} and to
specify the resource needs of each phase of the application. \two
\textbf{\emph{The coordinator}}, a hardware-based adaptive runtime system, 
uses the compiler annotations to dynamically allocate/deallocate resources for
each thread at phase boundaries. The coordinator plays the key role of
continuously controlling the extent of the oversubscription at each phase boundary.
\three \textbf{\emph{Hardware
virtualization support}} includes a mapping table for each resource to
locate each virtual resource in either the physically available on-chip
resources or the swap
space in main memory, and the machinery to swap resources between the physical
space and the swap space. 

Zorua has two key hardware components:
\emph{(i)}~the \emph{coordinator} that contains queues to buffer the
\emph{pending threads} and control logic to make oversubscription and 
resource management decisions, and
\emph{(ii)}~\emph{resource mapping tables}
to map each of the resources to their corresponding physical or swap spaces.
Our MICRO 2016 paper~\cite{zorua} provides the detailed implementation
of Zorua in Section~4. In particular, we describe several key issues, including
how
(1)~Zorua determines the amount of oversubscription for each resource (Section~4.4 of \cite{zorua}),
(2)~Zorua virtualizes each resource (Section~4.5 of \cite{zorua}), and
(3)~the compiler identifies each phase (Section~4.6 of \cite{zorua}).

\section{Results}
\label{sec:eval} 
In this section, we evaluate the effectiveness of \X in improving programming
ease, portability, and performance. Our detailed experimental methodology is
described in Section~5 of our MICRO 2016 paper~\cite{zorua}.
More results are provided in Section~6 of \cite{zorua}.

\subsection{Effect on Performance Variation and Cliffs}
\label{sec:eval:var}
\begin{figure}[h]
 \centering 
 \includegraphics[width=0.47\textwidth]{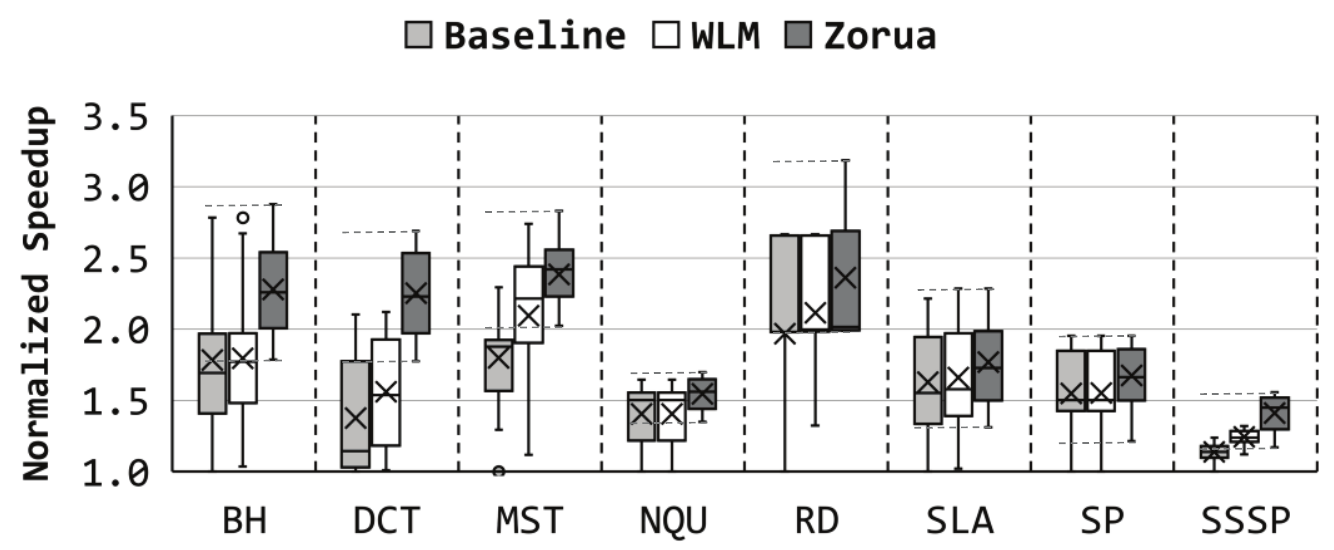}
\vspace{-2mm}
 \caption{Normalized performance distribution. Reproduced from~\cite{zorua}.}
 \label{fig:performance_range}
\end{figure}

We first examine how \X alleviates the high variation in performance by reducing
the impact of resource specifications on resource utilization.
Figure~\ref{fig:performance_range} summarizes the range in performance across a
wide range of resource specifications (indicating an undesirable dependence on
the specification), for the baseline architecture, WLM (which allocates
resources at the finer granularity of a warp~\cite{warp-level-divergence}), and Zorua
for a representative set of applications, using a Tukey box plot~\cite{mcgill1978variations}. The boxes in the box plot represent the range
between the first quartile (25\%) and the third quartile (75\%). 
The whiskers extending from the
boxes represent the maximum and minimum points of the distribution, or
1.5$\times$ the
length of the box, whichever is smaller. Any points that lie more than
1.5$\times$
the box length beyond the box are considered to be
outliers~\cite{mcgill1978variations}, and are plotted as
individual points. The line in the middle of the box represents the median,
while the ``X'' represents the average. 
We make two major
observations from Figure~\ref{fig:performance_range}.

First, we find that \X significantly reduces the \emph{performance range} across
all evaluated resource specifications. Averaged across all of our applications, the 
worst resource specification for Baseline achieves 96.6\% lower performance 
than the best performing resource specification. For WLM~\cite{warp-level-divergence},
this performance range reduces only slightly, to
88.3\%. With \X, the performance range drops significantly, to 48.2\%.
We see
drops in the performance range for \emph{all} applications except \emph{SSSP}. With \emph{SSSP}, the
range is already small to begin with (23.8\% in Baseline), and \X
exploits the dynamic underutilization, which improves performance but also adds a small amount of variation.

Second, while \X reduces the performance range, it also preserves or improves performance
of the best performing points. As we examine in more detail in
Section~\ref{sec:eval:perf}, the reduction in performance range occurs as
a result of improved performance mainly at the lower end of the distribution.

\begin{figure*}[ht!]
 \centering
 \begin{subfigure}[t]{0.297\linewidth}
 \centering
 \includegraphics[width=1.0\textwidth]{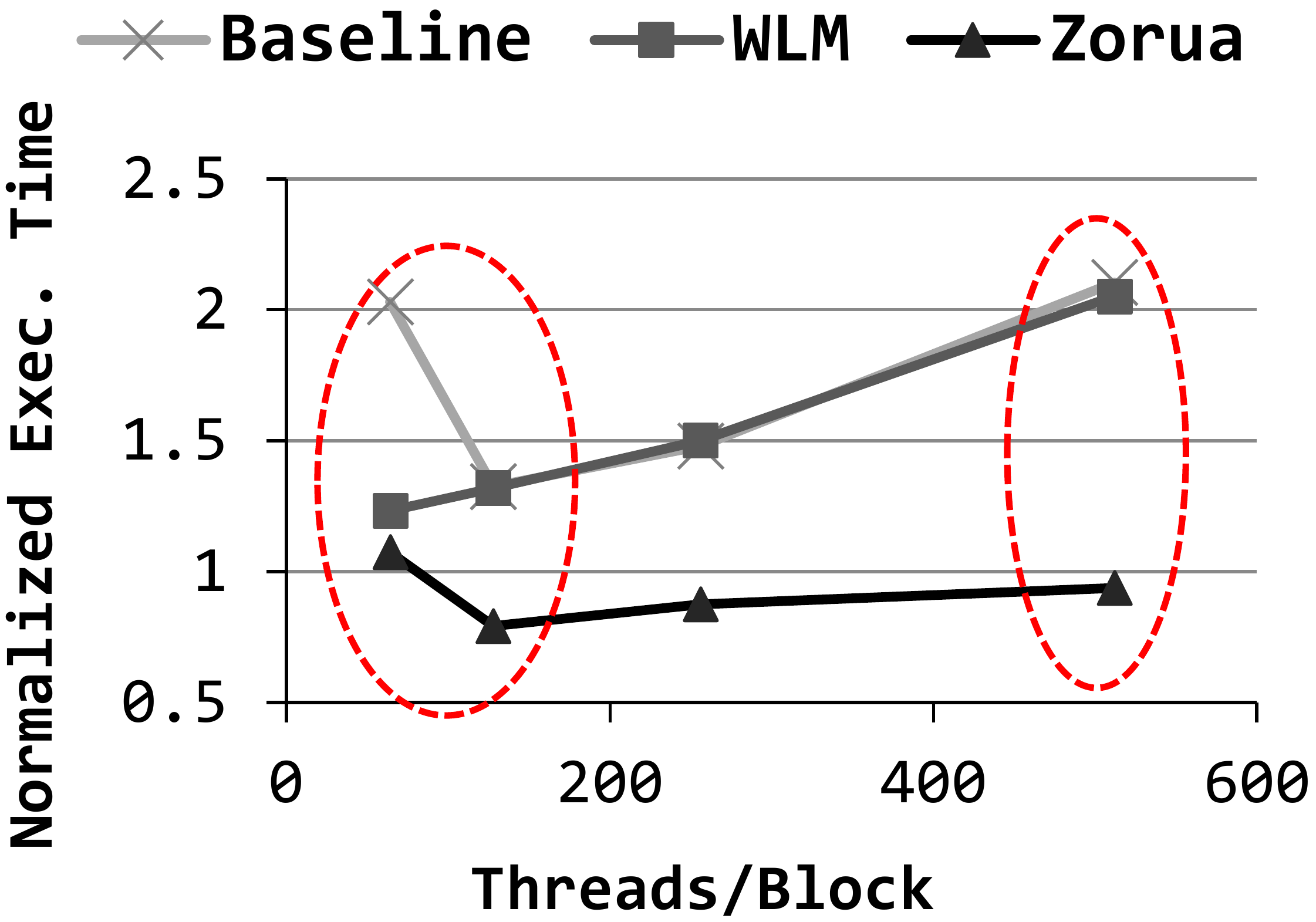}
\vspace{-4mm}
 \caption{\emph{DCT}} 
 \label{fig:performance_cliff_result_dct}
 \end{subfigure}\qquad
 \begin{subfigure}[t]{0.28\linewidth}
 \centering
 \includegraphics[width=1.0\textwidth]{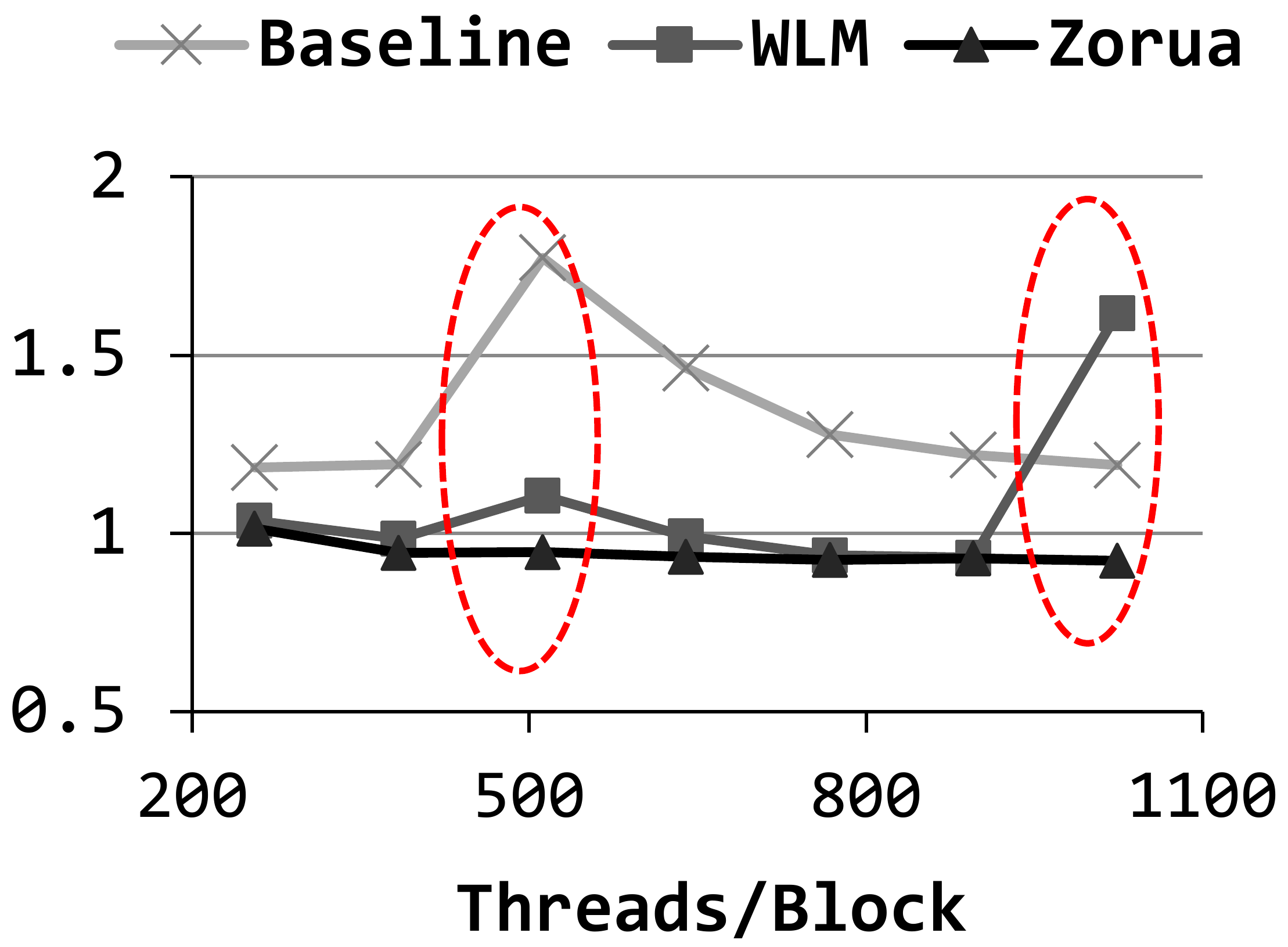} 
\vspace{-4mm}
 \caption{\emph{MST}}
 \label{fig:performance_cliff_result_mst}
 \end{subfigure}\qquad
 \begin{subfigure}[t]{0.28\linewidth}
 \centering
 \includegraphics[width=1.0\textwidth]{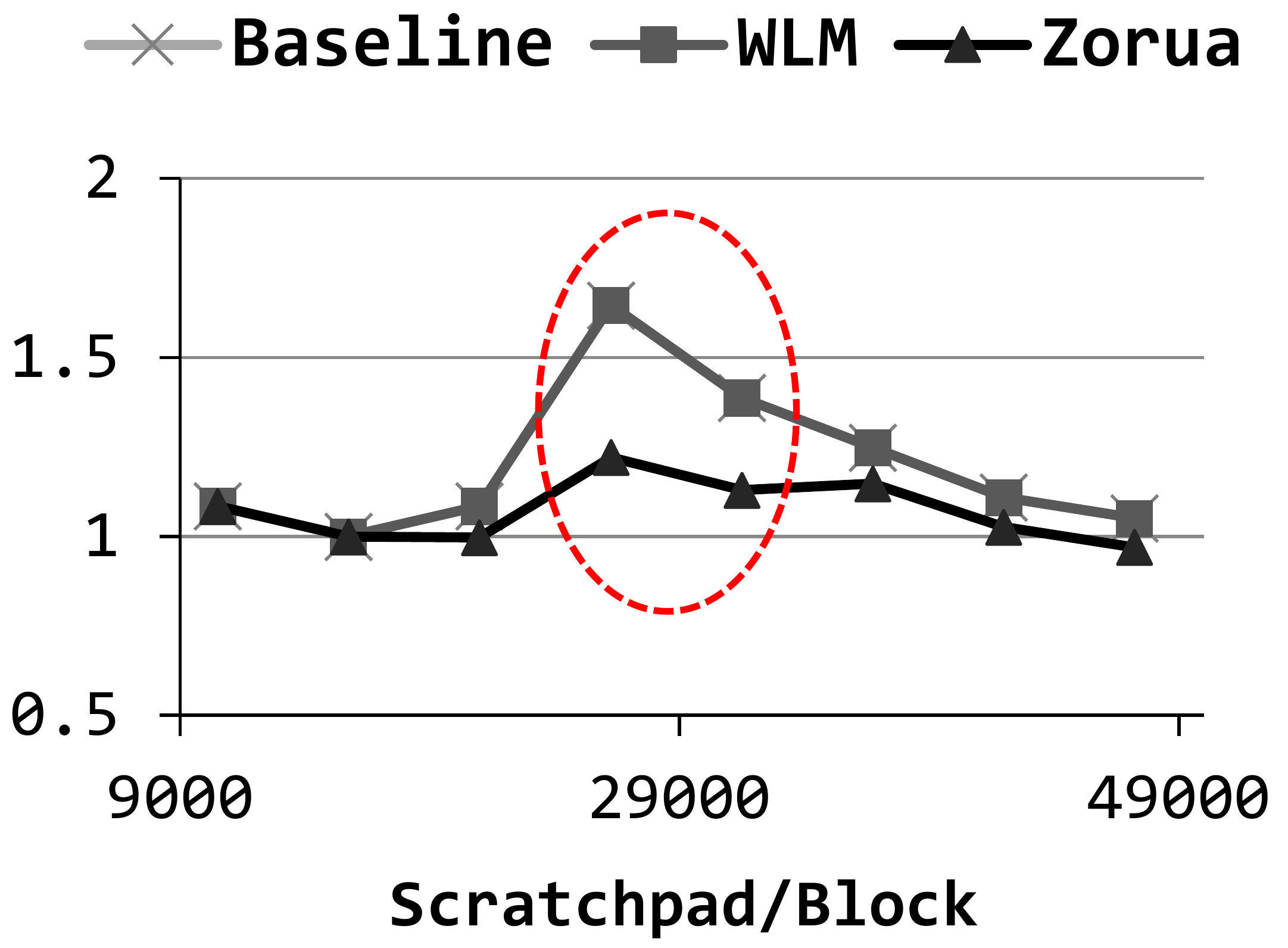}
\vspace{-4mm}
 \caption{\emph{NQU}}
 \label{fig:performance_cliff_result_nqu}
 \end{subfigure}
 \caption{Effect on performance cliffs. Reproduced from~\cite{zorua}.}
 \label{fig:performance_cliff_result}
\end{figure*}

To gain insight into how \X reduces the performance range and improves
performance for the worst performing points, we analyze how it reduces
performance cliffs. We study the tradeoff 
between resource specification and execution time for three representative 
applications: 
\emph{DCT} (Figure~\ref{fig:performance_cliff_result_dct}), \emph{MST}
(Figure~\ref{fig:performance_cliff_result_mst}), and
\emph{NQU} (Figure~\ref{fig:performance_cliff_result_nqu}).
For all three figures, we normalize execution time to the \emph{best} execution time
under Baseline. We make two observations from the figures.

First, \X successfully mitigates the performance cliffs that occur
in Baseline. For example, \emph{DCT} and \emph{MST} are both sensitive to the thread block size, as shown in 
Figures~\ref{fig:performance_cliff_result_dct}
and~\ref{fig:performance_cliff_result_mst}, respectively. We have circled the
locations at which cliffs exist in Baseline. Unlike Baseline, \X
maintains more steady execution times across the  
number of threads per block, 
employing 
oversubscription to overcome the loss in parallelism due to insufficient on-chip
resources. We see
similar results across all of our applications.

Second, we observe that while WLM~\cite{warp-level-divergence} can reduce some of the cliffs by mitigating
the impact of large block sizes, many cliffs still
exist under WLM (e.g., \emph{NQU} in Figure~\ref{fig:performance_cliff_result_nqu}).
This cliff in \emph{NQU} occurs as a result of insufficient scratchpad memory, which
cannot be handled by warp-level management. Similarly, the cliffs for \emph{MST}
(Figure~\ref{fig:performance_cliff_result_mst}) also persist with WLM because \emph{MST}
has a lot of barrier operations, and the additional warps scheduled by WLM
ultimately stall, waiting for other warps within the same block
to acquire resources. We find that, with oversubscription, \X is able to 
smooth out those cliffs that WLM
is unable to eliminate.

\subsection{Effect on Performance}
\label{sec:eval:perf}

As Figure~\ref{fig:performance_range} shows,
\X either retains or improves the best performing point for each application,
compared to the Baseline. \X improves the best performing point for each
application by 12.8\% on average, and by as much as 27.8\%
(for \emph{DCT}). This improvement comes from the improved parallelism obtained by exploiting the
dynamic underutilization of resources, which exists \emph{even for optimized
specifications}. Applications such as \emph{SP} and \emph{SLA} have little dynamic
underutilization, and hence do not show any performance improvement.
\emph{NQU} \emph{does}
have significant dynamic underutilization, but \X does not significantly improve the best
performing point as the overhead of oversubscription outweighs the benefit,
and \X dynamically chooses \emph{not} to oversubscribe. We conclude that even for
many specifications that are \emph{optimized} to fit the underlying hardware resources, 
\X is able to further improve performance. 

We also note that, in addition to reducing performance variation and
improving performance for optimized points, \X improves performance
by \emph{25.2\% on average} for all resource specifications across all evaluated
applications.

\subsection{Effect on Portability}
\label{sec:eval:port}

Performance cliffs often behave differently across different GPU architectures, and can
significantly shift the best performing resource specification point.
We study how \X can ease the burden of performance tuning if an application 
has been already tuned for one GPU model, and is later ported to another GPU. 
To understand this, we define a new
metric, \emph{porting performance loss}, that quantifies the performance impact
of porting an application without re-tuning it. To calculate this, we first
normalize the execution time of each specification point to the execution time
of the best performing specification point. We then pick a
source GPU architecture (i.e., the architecture that the GPU was tuned for) and
a target GPU architecture (i.e., the architecture that the code will run on),
and find the point-to-point drop in performance (when the code is executed on the target GPU) for all points whose
performance on the source GPU comes within 5\% of the performance at the
best performing specification point.\footnote{We include any point within 5\%
of the best performance as there are often multiple points close to the
best point, and the programmer may choose any of them.}

\begin{figure}[h]
 \centering 
 \includegraphics[width=0.48\textwidth]{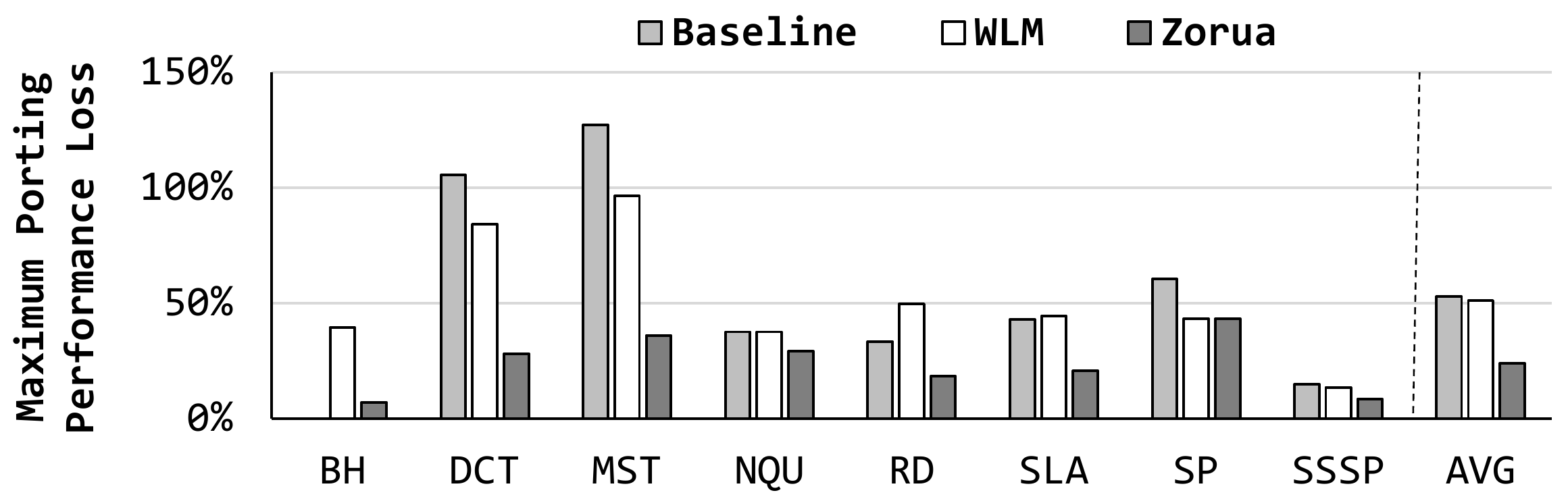}
 \caption{Maximum porting performance loss. Reproduced from~\cite{zorua}.}
 \label{fig:portability_result_overall}
\end{figure}

Figure~\ref{fig:portability_result_overall} shows the \emph{maximum} porting
performance loss for each application, across any two pairings of our three
simulated GPU architectures (NVIDIA Fermi, Kepler, and Maxwell). We find that \X greatly 
reduces the maximum porting performance loss that occurs under both Baseline
and WLM for all but one of our applications. On average, the maximum porting performance 
loss is 52.7\% for Baseline, 51.0\% for WLM, and only 23.9\% for \X.

Notably, \X delivers significant improvements in portability for applications 
that previously suffered greatly when ported to another GPU, such as \emph{DCT} and 
\emph{MST}. For both of these applications, the performance variation differs so much
between GPU architectures that, despite tuning the application on the source
GPU to be within 5\% of the best achievable performance, their performance on the target
GPU is often more than twice as slow as the best achievable performance on the target 
platform. \X significantly lowers this porting performance loss down to 28.1\% for \emph{DCT} and 36.1\% for \emph{MST}. We also observe
that for \emph{BH}, \X
actually slightly increases the porting performance loss with respect to the
Baseline. This is because for Baseline, there are only two points that perform within the 
5\% margin for our metric, whereas with \X, we have five points that fall in
that range. Despite this, the increase in porting performance loss for \emph{BH} is low, deviating
only 7.0\% from the best performance.

We conclude that \X enhances portability of applications by reducing the
impact of a change in the hardware resources for a given resource
specification. For applications that have 
already been tuned on one platform, \X significantly lowers the penalty of not 
re-tuning for another platform, allowing programmers to save development time.

\section{Related Work}
\label{sec:related}


To our knowledge, our MICRO 2016 paper~\cite{zorua} is the first work to propose a holistic framework to
decouple a GPU application's resource specification from its physical on-chip
resource allocation by virtualizing multiple on-chip resources. This
enables the illusion of more resources than what physically exists to the
programmer, while the hardware resources are managed at runtime by employing a
swap space (in main memory), transparently to the programmer. We design a new
hardware/software cooperative framework to effectively virtualize multiple
on-chip GPU resources in a controlled and coordinated manner, thus enabling many
benefits of virtualization in GPUs. 

We briefly discuss prior work related to different aspects of our
proposal: 
\One virtualization of resources,
\two improving
programming ease and portability,
and
\three more efficient management of
on-chip resources. 
 
\textbf{Virtualization of Resources.}
 \emph{Virtualization}~\cite{virtual-memory1,virtual-memory2,virtualization-1,virtualization-2}
 is a concept designed to provide the illusion, to the software and
 programmer, of more resources than what truly exists in physical
 hardware. It has been applied to the management of hardware
 resources in many different contexts
 ~\cite{virtual-memory1,virtual-memory2,virtualization-1,virtualization-2,vmware-osdi02,how-to-fake,pdp-10,ibm-360},
 with virtual memory~\cite{virtual-memory1, virtual-memory2, multics, fotheringham.cacm61}
 being one of the oldest forms of virtualization that is commonly
 used in high-performance processors today. Abstraction of hardware
 resources and use of a level of indirection in their management
 leads to many benefits, including improved utilization,
 programmability, portability, isolation, protection, sharing, and 
 oversubscription.

In this work, we apply the general principle of virtualization to the
management of multiple on-chip resources in modern
GPUs. Virtualization of on-chip resources offers the opportunity to
alleviate many different challenges in modern GPUs. However, in this
context, effectively adding a level of indirection introduces new
challenges, necessitating the design of a new virtualization
strategy. There are two key challenges. First, we need to dynamically determine
the \emph{extent} of the virtualization to reach an effective tradeoff
between improved parallelism due to oversubscription and
the latency/capacity overheads of swap space usage. Second,
we need to coordinate the virtualization of \emph{multiple} latency-critical
on-chip resources. To our knowledge, this is the first work to propose
a holistic software-hardware cooperative approach to virtualizing
multiple on-chip resources in a controlled and coordinated manner that
addresses these challenges, enabling the different benefits provided
by virtualization in modern GPUs.

Prior works propose to virtualize a specific on-chip resource for
specific benefits, mostly in the CPU context. For example, in CPUs,
the concept of virtualized registers was first used in the IBM
360~\cite{ibm-360} and DEC PDP-10~\cite{pdp-10} architectures to allow
logical registers to be mapped to either fast yet expensive physical
registers, or slow and cheap memory. More recent works~\cite{how-to-fake,cpu-virt-regs-1,cpu-virt-regs-2}, propose to virtualize registers to increase the effective
register file size to much larger register counts. This increases the
number of thread contexts that can be supported in a multi-threaded
processor~\cite{how-to-fake}, or reduces register spills and
fills~\cite{cpu-virt-regs-1,cpu-virt-regs-2}.\ignore{ Virtual Local
Stores~\cite{virtual-local-stores} is a scratchpad virtualization
mechanism to map the scratchpad inside the hardware-managed cache and
enable context-switching of the scratchpad state along with the rest
of the process state.} Other works propose to virtualize on-chip
resources in CPUs
(e.g.,~\cite{virtual-local-stores,spills-fills-kills,hierarchical-scheduling-windows,twolevel-hierarchical-registerfile,virtual-physical-registers-hpca98}). In
GPUs, Jeon et al.~\cite{virtual-register} propose to virtualize the
register file by dynamically allocating and deallocating physical
registers to enable more parallelism with smaller, more
power-efficient physical register files. 
Concurrent to this work, Yoon et al.~\cite{virtual-thread} propose an approach to virtualize thread slots
to increase thread-level parallelism. 
These works propose
specific virtualization mechanisms for a single resource for specific
benefits. None of these works provide a cohesive virtualization
mechanism for \emph{multiple} on-chip GPU resources in a
controlled and coordinated manner, which forms a key contribution of
our MICRO 2016 work. 

\textbf{Enhancing Programming Ease and Portability.}
There is a large body of work that aims to improve programmability and
portability of modern GPU applications using software tools, such as
auto-tuners~\cite{toward-autotuning,atune,maestro,parameter-profiler,autotuner1,autotuner-fft},
optimizing
compilers~\cite{g-adapt,optimizing-compiler1,parameter-selection,porple,optimizing-compiler2,sponge},
and high-level programming languages and
runtimes~\cite{cuda-lite,halide,hmpp,hicuda}. These tools tackle a multitude of
optimization challenges, and have been
demonstrated to be very effective in generating high-performance portable
code. They can also be used to tune the resource specification. 
However, there are several shortcomings in these approaches. First, these tools
often require profiling
runs~\cite{toward-autotuning,atune,maestro,porple,optimizing-compiler1,optimizing-compiler2}
on the GPU to determine the best performing resource specifications. These runs have to
be repeated for each new input set and GPU generation. Second, software-based approaches still require significant
programmer effort to write code in a manner that can be exploited by these
approaches
to optimize the resource specifications.\ignore{ For example, auto-tuners require
\emph{parameterization} of code, where the programmer is required to ensure
correctness of the program for any of the possible specification that an
auto-tuner optimizes. Optimizing compilers require programmers to write
kernels to ensure that each thread block is sized as small as possible for the
algorithm being implemented as the compiler has to conservatively preserve
synchronization primitives within a thread block. Some high-level languages~\cite{cuda-lite,halide,hmpp,hicuda} 
and compilers~\cite{g-adapt}
require annotations from the programmer or require the program to be written in
such a way that the algorithm is decoupled from potential optimization
schedules.} Third, selecting the best performing resource specifications statically using
software tools is a challenging task in virtualized
environments (e.g., cloud computing, data centers), where it is unclear which kernels
may be run together on the same SM or where it is not known, a priori, which GPU
generation the  application 
may execute on. Finally, software tools assume a fixed amount of available
resources. This leads to runtime underutilization due to static allocation of
resources, which cannot be addressed by these tools. 

In contrast, the programmability and portability benefits provided by \X
require no programmer effort in optimizing resource specifications. Furthermore,
these auto-tuners and compilers can be used in conjunction with \X to
further improve performance.

\textbf{Efficient Resource Management.}
Prior works aim to improve parallelism by increasing resource utilization
using
hardware-based~\cite{warp-level-divergence,shmem-multiplexing,unified-register,virtual-register,alternative-thread-block,register-mapping-patent,owl-asplos13,osp-isca13,largewarp,medic,rachata-isca,toggle-aware,decoupled-dma, usui.taco16}, software-based~\cite{shmem-multiplexing,stash,
asplos-sree,automatic-placement,onchip-allocation,enabling_coordinated,fine-grain-hotpar},
and hardware-software cooperative~\cite{mask, mosaic, caba, bis, acs,
marshaling, uba-joao-isca13, ltrf-sadrosadati-asplos18}
approaches.
%
Among these works, the closest to ours
are~\cite{virtual-register,virtual-thread}
(discussed earlier), \cite{shmem-multiplexing}, and \cite{warp-level-divergence}.
These approaches propose efficient techniques to
dynamically manage a single resource, and can be used along with \X to improve
resource efficiency further. 
Yang et
al.~\cite{shmem-multiplexing} aim to maximize utilization of the scratchpad with software
techniques, and by dynamically allocating/deallocating scratchpad memory. 
Xiang et al.~\cite{warp-level-divergence} propose
to improve resource utilization by scheduling threads at the finer granularity
of a warp rather than a thread block. This approach can help alleviate
performance cliffs, but not 
in the presence of synchronization or scratchpad
memory, nor does it address the dynamic underutilization within a thread during
runtime. We quantitatively compare to this approach in Section~\ref{sec:eval} and demonstrate
\X's benefits over it. 

Other works leverage resource
underutilization to improve energy efficiency~\cite{warped-register,energy-register,gebhart-hierarchical,compiler-register,virtual-register}
or perform other useful
work~\cite{caba,spareregister}. These works are
complementary to \X.


\section{Significance and Long-Term Impact}
In this section, we describe
the significance and long-term impact of our MICRO 2016 work, Zorua, by delineating its novelty,
what it can enable in future systems, and new
research directions that it triggers.

\subsection{Novelty}
\noindent{\textbullet}~This is the first work that takes a holistic approach to
decoupling a GPU application's resource specification from its physical on-chip
resource allocation via the use of virtualization. We develop a comprehensive
virtualization framework that provides \emph{controlled} and \emph{coordinated}
virtualization of \emph{multiple} on-chip resources to maximize the
effectiveness
of virtualization. 

\vspace{5pt}%
\noindent{\textbullet}~Making GPUs easy to program is critical for their widespread use, and
also to achieve the high performance promised by the massively parallel
architecture. A key limiting factor in GPU programming today is the
burden placed on the programmer in finding a hardware resource
specification that achieves very high
performance. This is the first work to ease that burden without compromising
performance by virtualizing
the major hardware resources programmers are required to manage today.%

\vspace{5pt}%
\noindent{\textbullet}~Portability across GPU architectures is vital in environments such as
cloud computing and data centers to achieve predictably good
performance, \emph{irrespective} of the GPU generation the application is executing on.
This is the first work to tackle the portability challenges that arise from the
\highlight{programmer's management} of the fixed on-chip resources with a holistic resource virtualization
strategy. 
\subsection{What Zorua Can Enable in Future Systems}
GPUs have emerged as the dominant massively parallel GPU architecture, used as the
platform of choice for a wide range of parallel applications from machine
learning to scientific simulation. However, there are a number of key challenges
that limit the adoption of GPUs across broader classes of
applications and environments, e.g., data centers, cloud computing, etc. Programmability and
portability of GPU applications are two such challenges. But future GPUs will
need to address several other challenges before truly becoming first-class
compute engines. As we describe below, we believe that our work can help address
some of these other challenges.
\vspace{1mm}

\textbf{Multiprogramming in Virtualized Environments.} 
Zorua lends itself to easily
addressing two key challenges in enabling 
multiprogramming in virtualized environments today:

\emph{Fine-grained resource sharing across kernels:}
\X manages the different
resources independently and at a fine granularity, using a dynamic runtime system. 
Hence, Zorua can be extended to support fine-grained sharing and partitioning
of resources across multiple kernels to enable efficient multiprogramming in GPUs. 
Zorua enables better resource utilization in these multiprogrammed environments, while providing the ability to control the partitioning of
resources at runtime to provide QoS, fairness, etc., by leveraging the hardware
runtime system. \highlight{Zorua can work synergistically with systems
such as Mosaic~\cite{mosaic} and MASK~\cite{mask}, which enable efficient
memory virtualization techniques for GPUs, to enable true full-system multi-kernel
execution.}

\emph{Preemptive multitasking:} Another key challenge in enabling true multiprogramming in GPUs is
enabling rapid preemption of
kernels~\cite{isca-2014-preemptive,simultaneous-sharing,chimera}. Context
switching on GPUs incurs a very high latency and overhead, as a result of the large amount of
register file/scratchpad state that needs to be saved 
before a new kernel can be executed. 
\X
enables fine-grained management and virtualization of on-chip resources. It can be naturally extended to enable quick
preemption of a task via intelligent management of the swap space and the
mapping tables. It can also work synergistically with CABA~\cite{caba},
framework for assist warp execution in GPUs, to provide flexible and efficient
support for multitasking and context switching. 

\textbf{Support for Other Parallel Programming Paradigms.}
The fixed static resource allocation for each thread in
modern GPU architectures requires statically dictating the
parallelism for the program throughout its execution. Other forms
of parallel execution that are \emph{dynamic} (e.g., CILK~\cite{cilk}) require more flexible allocation
of resources at runtime, and are hence more challenging to enable on GPUs.
Examples of this include \emph{nested parallelism}~\cite{nested}, where a kernel can dynamically spawn new kernels or
thread blocks, and \emph{helper threads}~\cite{caba} to utilize idle resource at runtime to
perform different optimizations or background tasks in parallel. Zorua makes it
easy to enable these paradigms by providing on-demand dynamic allocation of
resources. 

\textbf{Energy Efficiency, Scalability, and Reliability.}
To support massive parallelism, on-chip resources are a precious
and critical resource. However, these resources \emph{cannot} grow arbitrarily large as
GPUs continue to be area-limited and on-chip memory tends to be extremely power
hungry and area
intensive~\cite{energy-register,virtual-register,compiler-register,warped-register,virtual-thread,ltrf-sadrosadati-asplos18},
which are trends we believe will become increasingly important for the
foreseeable future.
Furthermore, complex thread schedulers that can select a thread for execution
from an increasingly large thread pool are required. \X enables using smaller
register files, scratchpad memory and less complex or fewer thread schedulers to
save power and area while still retaining or improving parallelism.
The indirection offered by \X, along with the dynamic
management of resources, could also enable better reliability. The
virtualization framework trivially allows portions of a resource that contain hard or soft
faults to be remapped to other portions of the resource that do not contain faults, 
or to spare structures, thereby increasing the error tolerance of these resources.

\subsection{New Research Directions Zorua Enables}
Zorua opens up several new avenues for more research, which we briefly discuss
here.

\textbf{Flexible Programming Models for GPUs and Heterogeneous
Systems.}
By providing a flexible but dynamically controlled view of the on-chip hardware resources, Zorua changes the abstraction of the on-chip resources
that is offered to the programmer and software. This offers the opportunity to
rethink resource management in GPUs from the ground up. One could envision more powerful resource
allocation and better programmability with programming models that do
\emph{not} require static
resource specification, leaving the compiler/runtime system and the underlying
virtualized framework to \highlight{\emph{completely}
handle \emph{all}} forms of on-chip resource
allocation, unconstrained by the fixed physical resources in a specific
GPU, entirely at runtime. This is especially
significant in future systems that are likely to support a wide range of
compute engines and accelerators, making it important to be able to write
high-level code that can be partitioned easily, efficiently, and at a
fine granularity across any \highlight{set of accelerators}, without statically tuning any code segment to run efficiently on the GPU.

\textbf{Virtualization-Aware Compilation and Auto-Tuning.} Zorua changes the
contract between the hardware and software to provide a more powerful
resource abstraction (in the software) that is
\emph{flexible and dynamic}, by pushing some more functionality to the hardware,
which can more easily react to runtime resource requirements of the program. We can re-imagine compilers
 and auto-tuners to be more intelligent, leveraging this new abstraction and,
 hence the virtualization, to
 deliver more efficient
and high-performing code optimizations 
that are \highlight{\emph{not} possible with the \emph{fixed} and
\emph{static}}
abstractions of today. They could, for example, \emph{leverage} the
oversubscription and dynamic management that Zorua provides to tune the code to
more aggressively use resources. 

\textbf{Support for System-Level Tasks on GPUs.}
As GPUs become increasingly general purpose, a key requirement is better
integration with the CPU operating system, and with complex distributed software
systems such as those employed for large-scale distributed machine
learning~\cite{tensorflow, gaia} or
graph processing\highlight{~\cite{graphlab, tesseract,pim-enabled}}. If GPUs are architected to be
first-class compute engines,
rather than the slave devices they are today, they can be programmed and utilized in the
same manner as a modern CPU. 
This integration requires the GPU execution model to support system-level
tasks like interrupts, exceptions, etc. and more
generally provide support for access to distributed file systems,
disk I/O, or network communication. Support for these tasks and
execution models require dynamic provisioning of resources for execution of
system-level code. Zorua provides a building block to enable this.

\textbf{Applicability to General Resource Management in Accelerators.} 
Zorua uses a program \emph{phase} as the granularity for managing resources.
This allows handling resources across phases
\emph{dynamically}, while leveraging \emph{static} information regarding resource requirements
from the software by inserting annotations at phase boundaries. Future work could potentially investigate the applicability
of the same approach to manage
resources and parallelism in \emph{other} accelerators (\highlight{e.g.,
processing-in-memory accelerators~\cite{pim-enabled, tesseract, tom,
impica,googlepim-asplos18, shaw1981non, boroumand2016pim, kim.bmc18,
ambit, guo-wondp14, stone1970logic, zhang-2014, kogge.iccp94, patterson.ieeemicro97} or direct-memory access engines~\cite{rowclone,decoupled-dma, chang.hpca16}}) that require efficient dynamic management of large
amounts of particular critical resources.

\section{Conclusion}

We propose \X, a new framework that decouples the application resource
specification from the allocation in the physical hardware resources (i.e.,
registers, scratchpad memory, and thread slots) in GPUs. \X
encompasses a holistic virtualization strategy to effectively
virtualize multiple latency-critical on-chip resources in a
controlled and coordinated manner. We demonstrate that by providing
the illusion of more resources than physically available, via dynamic
management of resources and the judicious use of a swap space in
main memory, \X enhances \One \emph{programming ease} (by reducing
the performance penalty of suboptimal resource specification), \two
\emph{portability} (by reducing the impact of different hardware
configurations), and \three \emph{performance} for code with an
optimized resource specification (by leveraging dynamic
underutilization of resources). We conclude that \X is an
effective, holistic virtualization framework for GPUs. We believe
that the indirection provided by {\X}'s virtualization mechanism
makes it a generic framework that can address other challenges in
modern GPUs. For example, \X can enable fine-grained resource
sharing and partitioning among multiple kernels/applications, as well as 
low-latency preemption of GPU programs. We hope that future work
explores these promising directions, building on the insights and
the framework developed in our MICRO 2016 paper.

\ignore{We conclude that by decoupling the programmer's view of the resources from what
is physically available, \X enhances
programming ease, portability, and performance of GPU applications while paving the
way for many other use cases that can leverage the fluid view of resources.}

\section*{Acknowledgments}

We thank the reviewers and our shepherd for their valuable suggestions. We thank the
members of the SAFARI group for their feedback and the stimulating research
environment they provide. Special thanks to Vivek Seshadri, Kathryn McKinley,
Steve Keckler, Evgeny Bolotin, and Mike O'Connor for their feedback during
various stages of this project. We acknowledge the support of our
industrial partners: Facebook, Google, IBM, Intel, Microsoft, NVIDIA,
Qualcomm, Samsung, and VMware. This research was partially supported
by NSF (grant 1409723), the Intel Science and Technology Center for
Cloud Computing, and the Semiconductor Research Corporation.

\bibliographystyle{IEEEtranS}
\bibliography{paper,gpu}

\end{document}